\documentclass{PoS}

\title{The CAST experiment: status and perspectives}

\ShortTitle{The CAST experiment: status and perspectives}
\author{\speaker{F.~J.~Iguaz}$^{1}$ on behalf of the CAST Collaboration\\
\llap{$^{1}$} IRFU, Centre d'\'Etudes Nucl\'eaires de Saclay (CEA-Saclay), Gif-sur-Yvette, France\\
E-mail: \email{franciscojose.iguazgutierrez@cea.fr}
}

\abstract{The status of the solar axion search with the CERN Axion Solar Telescope (CAST) will be discussed. Results from the first part of CAST phase II where the magnet bores were filled with $^4$He gas at variable pressure in order to scan m$_a$ up to 0.4 eV will be presented. From the absence of excess X-rays when the magnet was pointing to the Sun, we set a typical upper limit on the axion-photon coupling of g$_{a\gamma} < 2.17 \times 10^{−10}$ GeV$^{-1}$ at 95\% CL for m$_a \lesssim$ 0.4 eV, the exact result depending on the pressure setting. Our search for axions with masses up to about 1.2 eV using $^3$He as a buffer gas is in progress in the second part of CAST phase II. Expectations for sensibilities will be given. Near future perspectives as well as more long term options for a new helioscope experiment will be evoked.}

\FullConference{Identification of Dark Matter 2010-IDM2010\\
		July 26-30, 2010\\
		Montpellier France}

\begin{document}
\section{Introduction}
Quantum Chromodynamics (QCD) is expected to violate CP-symmetry but this effect has not been observed by any experiment. To explain this apparent conservation, R. Peccei and H. Quinn introduce in 1977 one additional symmetry \cite{HPeccei177}. As pointed out by S. Weinber and F. Wilczek in 1978, the broken of this symmetry at a yet unknown scale (f$_a$) gives rise to a Goldstone boson called \emph{axion} \cite{Weinberg78}. This particle is especially appealing since in addition to solving the strong CP-problem, the axion would also be an excellent candidate for Dark Matter due to its expected physical properties.

\medskip
Since the breaking scale f$_a$ is not a priori determined, the axion mass is initially unknown. Several constraints from astrophysics and cosmology have been applied to prove or rule out the existence of the axion. These limits can narrow the axion mass to a window between some $\mu$eV and 1 eV. There are also several experiments which have attempted to detect axions. Most of them use the so-called Primakoff effect \cite{HPrimakoff51}. It consists in the coupling of an axion to a virtual photon provided by a transverse magnetic field, being transformed into a real photon that carries the energy and the momentum of the original axion. From these experiments, one type are helioscopes \cite{KBibber87}, which try to detect the axions generated in the inner solar core. The Sun could generate axions with a mean energy of 4.2 keV, which can be reconverted to photons at Earth with the use of a transverse magnetic field. The resulting photons can be detected with conventional X-ray detectors.

\section{The CAST experiment}
The CERN Axion Solar Telescope (CAST) uses one of the prototypes of a superconducting LHC dipole magnet. It provides a magnetic field of up to 9 T transversal to the direction of axion propagation. The magnet is able to follow the Sun twice a day during the sunset and sunrise for a total of about 3 hours per day. During the remaining time, i.e, when the CAST magnet is not aligned witht the solar core, background data is taken.

\medskip
Four X-ray detectors are installed at both ends of the 10 m long magnet in order to search for photons from Primakoff conversion. Two Micromegas detectors \cite{PAbbon07} follow the sunset. They have replaced the formerly used Time Projection Chamber \cite{DAutiero07} and show a better performance in terms of background discrimination than the former detector. The sunrise is covered by another Micromegas detector and a combination of a X-ray mirror optics with a Charge Coupled Device (CCD) \cite{MKuster07}. The mirror focus the X-rays to a small spot on the CCD chip, improving the signal-to-background ratio and the sensitivity of the detector.

\section{CAST updates and recent results}
The experiment finished its first phase with vacuum by 2004 after having taken two years of data. As no significant signal above background was observed when following the Sun, an upper limit on the axion-to-photon coupling of $g_{a\gamma} < 8.8 \times 10^{-11}$ GeV$^{-1}$ (95\% CL) for axion masses $m_a \lesssim 0.02$ eV was set \cite{KZioutas05}. For extending the axion mass range, CAST was upgraded with a sophisticated gas system and novel cold windows for operation with helium gas at various pressures inside the magnetic field region. This gas would restore the coherent conversion probability from axions to photons for a narrow axion mass range.

\begin{figure}[htb!]
\centering
\includegraphics[width=.6\textwidth]{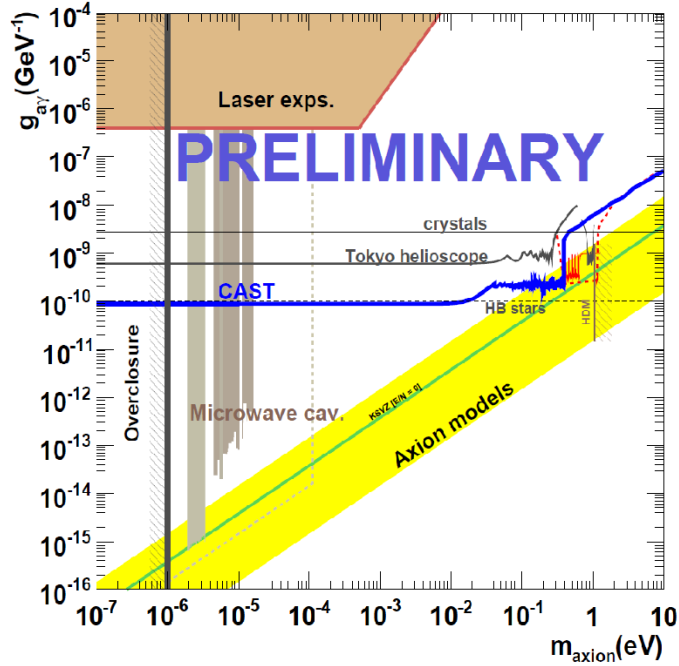}
\caption{The blue line is exclusion plot of the axion-to-photon coupling constant at 95\%~CL for all data obtained in Phase I and Phase II with $^4$He and $^3$He gas at CAST. The red line shows our prospects for the rest of the $^3$He phase. The latest limit set by the Tokyo helioscope Sumico has also been included \cite{YInoue08}.  The yellow band represents the usual axion theoretical models and the green solid line corresponds to the case of the KSVZ model with $E/N = 0$. Limits from laser, microwave and undeground detectors in axions searches have been also included \cite{ZAhmed09}.}
\label{Exclusion}
\end{figure}

\medskip
During 2005 and 2006 the magnet was filled with $^4$He gas and axion masses up to 0.39 eV were investigated by measuring a total of 160 presure steps between 0.08 mbar and 13.4 mbar. A typical upper limit on the axion-to-photon coupling of $g_{a\gamma} < 2.2 \times 10^{-10}$ GeV$^{-1}$ (95\% CL) for axion masses $m_a \lesssim 0.39$ eV was extracted \cite{EArik09}. The exact value of this limit depends on the considered pressure limit and the final results are shown in figure \ref{Exclusion} (left).

\begin{figure}[htb!]
\centering
\includegraphics[width=.6\textwidth]{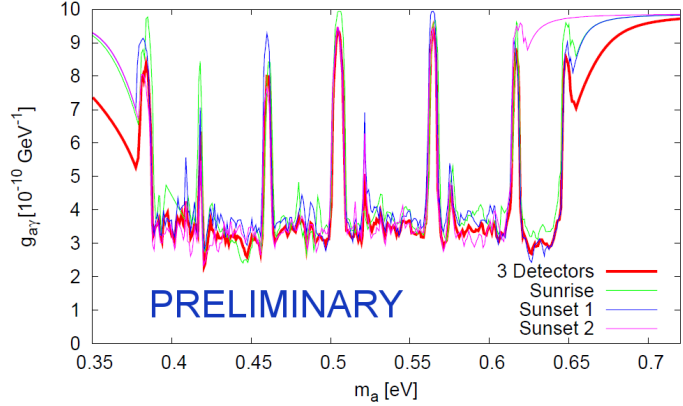}
\caption{The preliminary results obtained with $^3$He gas during 2008. The upper limit takes into account three of the four CAST detectors and covers axion masses from 0.39 to 0.65 eV.}
\label{Exclusion2}
\end{figure}

\medskip
Since the year 2008, CAST is taking data with $^3$He in the magnet bore. At the moment of writing this proceeding, nearly 720 pressure steps have already been measured, equivalent to a gas pressure of 80 mbar and an axion mass of 0.93 eV. In figure \ref{Exclusion2}, a preliminary exclusion plot for the $^3$He data acquired during 2008 is shown. It includes data from three out of the four detectors and covers axion masses between 0.39 and 0.65 eV. The data of the remaining detector is being analyzed and the final combined limit is expected soon.

\medskip
Other non-standard axion scenarios to which CAST would be also sensitive have also been studied. The data taken by the TPC in the phase I was reanalyzed to search for 14.4 keV axions coming from M1 transitions in the Sun \cite{Andriamonje09}. In addition, a calorimeter took data during the first phase to find MeV lines from high energy axion conversion \cite{Andriamonje10}. Finally, few days of data were carried out with a visible detector coupled to one end of the magnet, looking for axions in the visible range, possibly produced at the surface of the Sun. A preliminary result was recently released \cite{GCantatore08}. The detector has been now installed as a permanent detector in combination with the CAST sunrise micromegas detector.

\section{Future and outlook of CAST}
The collaboration is studying the extension of the experiment in the near-term in two different aspects. On one hand, other possible search candidates are being considered like standard-QCD axions, chameleons \cite{Brax10}, paraphotons and any other WISPs (Weakly Interacting Slim Particles). On the other hand, CAST is working in the development of the the next-generation of axion helioscope (NGAH), which could increase the actual sensitivity to $10^{-11}$ GeV$^{-1}$, i.e., one order of magnitude \cite{TPapaevangelou09}. We must note that the magnet features (length, field and aperture) have the main importance in the sensitiviy of an helioscope experiment, as this variable can be expressed as
\begin{equation}
g_{a\gamma} \propto \left(B \ L\right)^{-1/2} \ A^{-1/4} \ \left(\frac{b}{t}\right)^{1/8}
\label{eq:sens}
\end{equation}
where $B$ is the magnetic field, $L$ and $A$ are the lenght and aperture of the magnet, $b$ is the background level and $t$ is the time of exposure.

\begin{figure}[htb!]
\centering
\includegraphics[width=.35\textwidth]{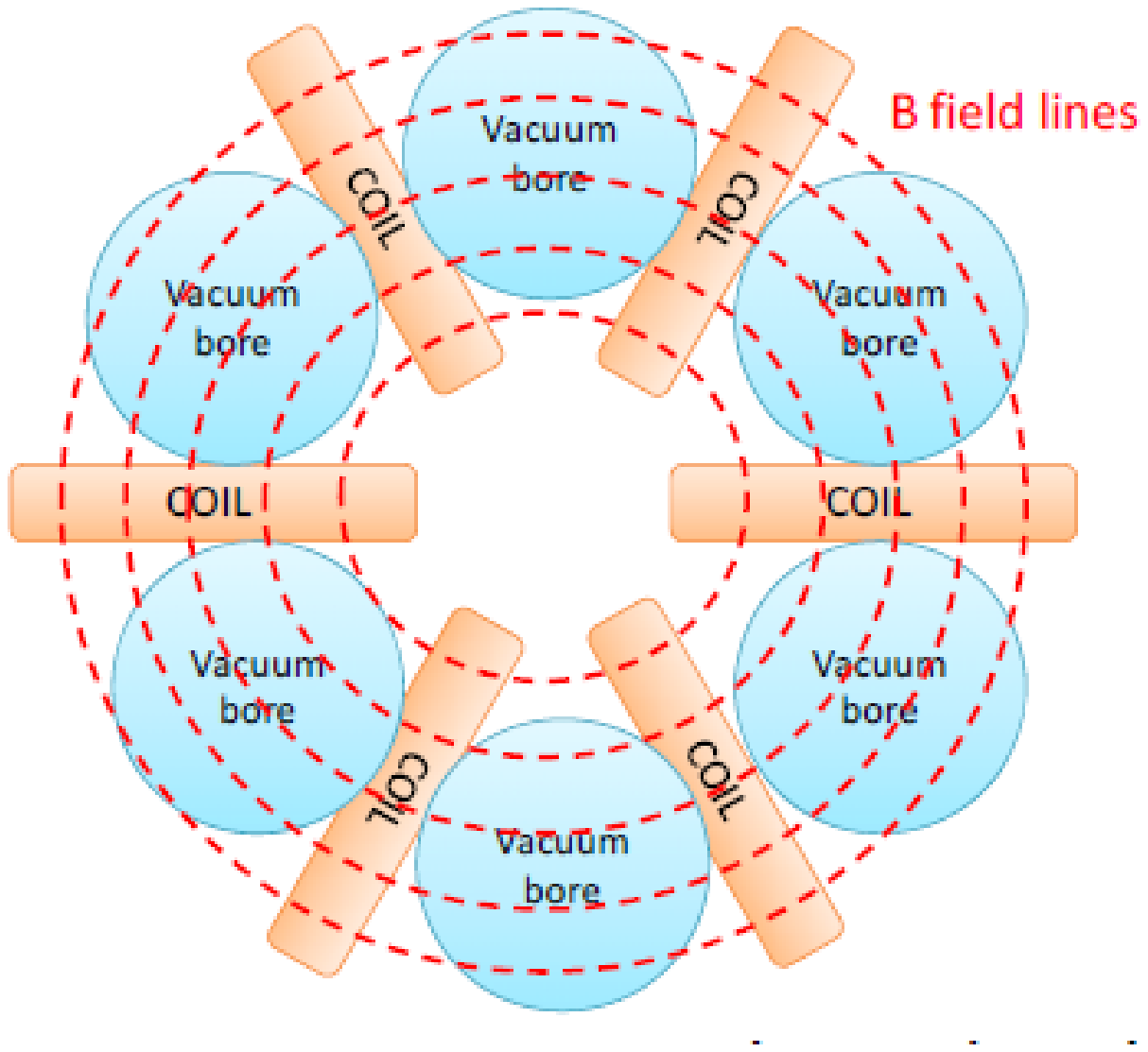}
\hspace{10.0mm}
\includegraphics[width=.45\textwidth]{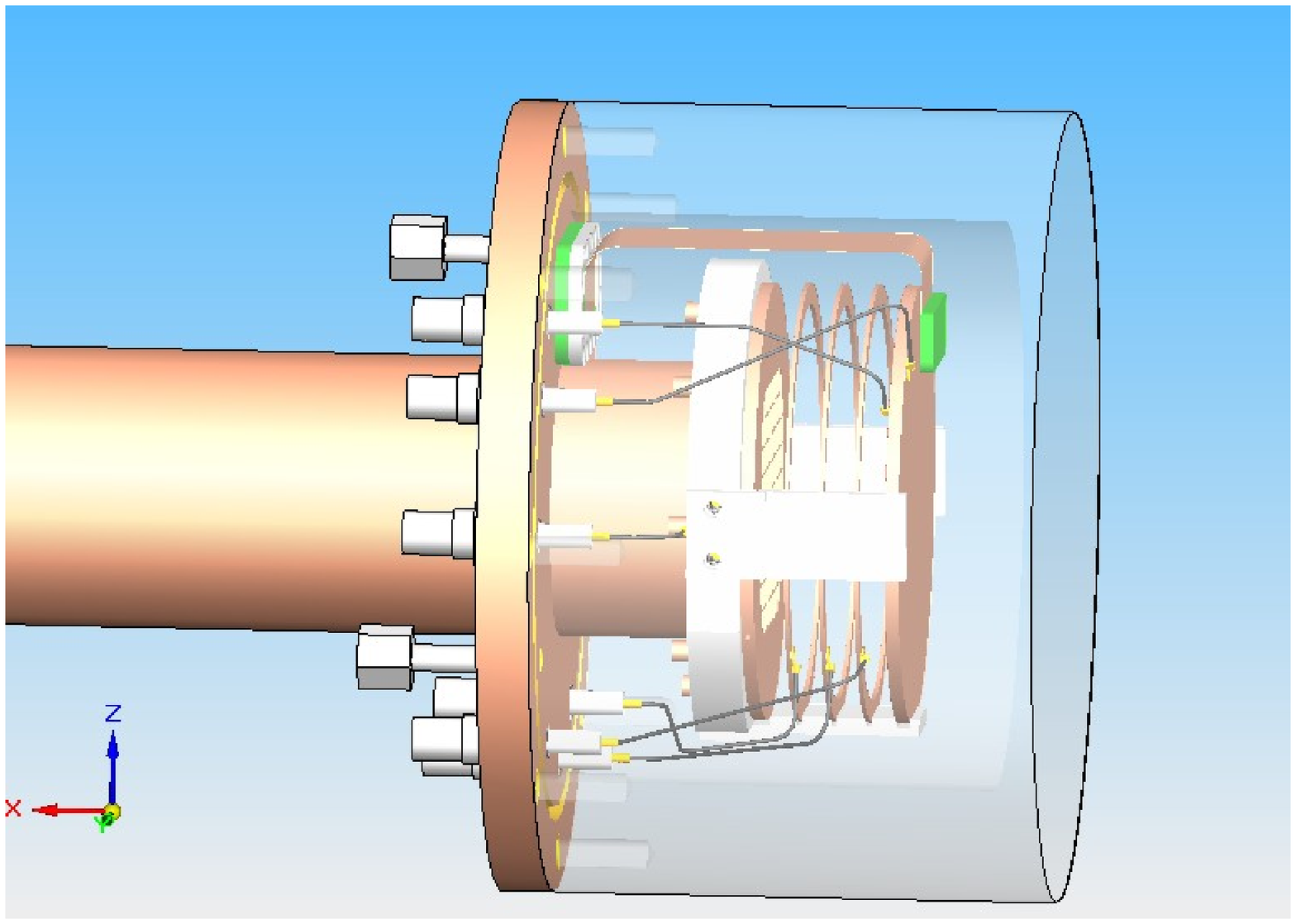}
\caption{Left: The toroidal magnet proposed in \cite{LWalckiers10}. Right: Design of the new Micromegas detector proposed by J.P.~Mols. It is composed of a lead cylindrical vessel (transparent) and a copper plate (pink) which keep the vessel gas-tight. Both materials show good radio-purity properties and shielding performances. Strips signals are taken out by a plain cable and a feedthrough, and then connected to the electronics.}
\label{MagnetDet}
\end{figure}

\medskip
The most promising configuration is the toroidal magnet \cite{LWalckiers10}, shown in figure \ref{MagnetDet} (left), which consists in several coils separated by vacuum bores. The main improvements of this design compared to the actual CAST dipole are its big aperture (near 1 m$^2$) and its lighter construction.

\medskip
In parallel, efforts are made in the development of high efficiency focusing devices and new electronics for the X-ray detectors, aiming to achieve very low background levels. For this purpouse, a new Micromegas detector for CAST is being designed, optimised for the next generation of helioscopes. A possible design is shown in figure \ref{MagnetDet} (right). Low radioactive materials will be used in the manufacturing of the microbulk detector and the air-tightness of the setup will be also improved, as there are indications of its main role in background. The use of the T2K electronics \cite{PBaron08} for increasing the rejection capabilities of the detector is also being studied.

\begin{figure}[htb!]
\centering
\includegraphics[width=.53\textwidth]{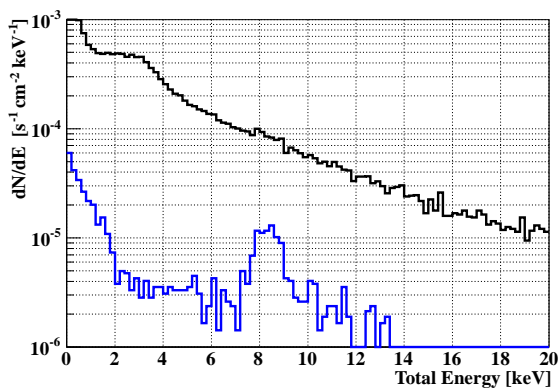}
\hspace{1mm}
\includegraphics[width=.45\textwidth]{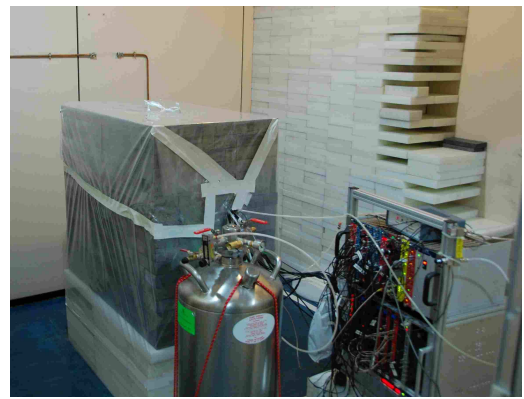}
\caption{Left: Background energy spectrum of a CAST Micromegas detector shielded with 5 cm of lead. The black line considers all the events and the blue one only those selected by the discrimination algorithms. Right: A Micromegas CAST detector installed inside a 20 cm-thick lead castle at the Canfranc Underground Laboratory. A constant flux of nitrogen keep the inside atmosphere clean of radon.}
\label{SimCanf}
\end{figure}

\medskip
Since the year 2008, several ultra-low background periods have been observed in different Micromegas detectors. During those periods, the background level suddenly decreased one order of magnitude from its mean value, around $4 \times 10^{-6}$ s$^{-1}$ keV$^{-1}$ cm$^{-2}$ in the range between 2 and 7 keV \cite{ATomas10} as shown in figure \ref{SimCanf} (left). This effect was found to be correlated with very low concentrations of radon. Two activities have been started to confirm this hypothesis. On one hand, a detailed plan of simulations is carried out to understand the contribution of environmental and internal contaminations in the final background of the Micromegas detector. On the other hand, a Micromegas CAST detector has been installed in the Canfranc Underground Laboratory, shielded by lead and copper and kept in a radon-clean atmosphere, as shown in figure \ref{SimCanf} (right).

\section{Conclusions}
The CAST experiment has established the most stringent experimental limit on the axion coupling constant over a wide range of masses, exceeding astrophysical constraints. The $^4$He phase has allowed to enter in the region favoured by axion models. From the absence of X-ray excess when the magnet was pointing to the Sun, a preliminary upper limit on the axion-photon coupling of $g_{a\gamma} \leq 2.22 \times 10^{-10}$ GeV$^{-1}$ at 95\% CL has been set for axion masses $m_a \lesssim 0.4$ eV. The exact result depends on the pressure setting. With the $^3$He phase, CAST is currently extending its axion search even further into the unexplored regions of the favored axion models. A preliminary result of 2008 data has been presented.

\medskip
The collaboration is taking part in the development of the next generation of helioscopes and of high efficiency focusing devices and new electronics for the future X-ray detectors. Those efforts try to extend the sensitivity of the experiment to the order of $10^{-11}$ GeV$^{-1}$, leading to explore a large part of the QCD favoured model region in combination with the Microwave Cavity experiments (ADMX) in the next years.


\begin{thebibliography}{99}
\bibitem{HPeccei177} R.~Peccei, H.~Quinn, \emph{Phys. Rev. Letters} {\bf 38} (1977) 1440.
\bibitem{Weinberg78} S.~Weinberg, \emph{Phys. Rev. Letters} {\bf 40} (1978) 223; F.~Wilczek, \emph{Phys. Rev. Letters} {\bf 40} (1978) 279.
\bibitem{HPrimakoff51} H.~Primakoff, \emph{Phys. Rev. Letters} {\bf 81} (1951) 899; P.~Sikivie, \emph{Phys. Rev. Letters} {\bf 51} (1983) 1415 (Erratum ibid. {\bf 52} (1984) 695).
\bibitem{KBibber87} K.~van~Bibber {\it et al.}, \emph{Phys. Rev. Letters} {\bf 59} (1987) 759.
\bibitem{KZioutas05} K.~Zioutas {\it et al.}, \emph{Phys. Rev. Letters} {\bf 94} (2005) 121301; S.~Andriamonje {\it et al.}, \emph{JCAP} {\bf 4} (2007) 010.
\bibitem{PAbbon07} P.~Abbon {\it et al.}, \emph{NJP} {\bf 9} (2007) 170.
\bibitem{DAutiero07} D.~Autiero {\it et al.}, \emph{NJP} {\bf 9} (2007) 171.
\bibitem{MKuster07} M.~Kuster, \emph{NJP} {\bf 9} (2007) 169.
\bibitem{EArik09} E.~Arik {\it et al.}, \emph{JCAP} {\bf 02} (2009) 008.
\bibitem{YInoue08} Y.~Inoue {\it et al.}, \emph{Phys. Lett. B} {\bf 668} (2008) 93.
\bibitem{ZAhmed09} Z. Ahmed {\it et al.}, \emph{Phys. Rev. Lett.} {\bf 103} (2009) 141802, \emph{arXiv}:0902.4693 [hep-ex].
\bibitem{Andriamonje09} S.~Andriamonje {\it et al.}, \emph{JACP} {\bf 0912} (2009) 002.
\bibitem{Andriamonje10} S.~Andriamonje {\it et al.}, \emph{JACP} {\bf 1003} (2010) 032.
\bibitem{GCantatore08} G.~Cantatore {\it et al.}, \emph{arXiv}:0809.4581 [hep-ex] (2008).
\bibitem{Brax10} P.~Brax and K.~Zioutas, \emph{Phys. Rev. D} {\bf 82} (2010) 043007, \emph{arXiv}:1004.1846 [astro-ph].
\bibitem{TPapaevangelou09} T.~Papaevangelou {\it et al.}, \emph{New opportunities in the Physics Landscape at CERN}, May 2009.
\bibitem{LWalckiers10} Suggested by Louis Walckiers / CERN.
\bibitem{PBaron08} P. Baron {\it et al.}, \emph{IEEE Nucl. Sci. Symp. Conf. Rec.} {\bf 55} (2008) 1744; D. Calvet {\it et al.}, \emph{16th IEEE NPSS real time conference}, 2009, Beijing (China).
\bibitem{ATomas10} A.~Tomas {\it et al.}, Proceedings of the conference EXRS2010, Coimbra (Portugal).
\end{thebibliography}
\end{document}